\documentclass[%
reprint,amsmath,amssymb]{revtex4}

\usepackage{graphicx}% Include figure files
\usepackage{epstopdf}
\usepackage[colorlinks=true,linkcolor=blue,citecolor=blue]{hyperref}
\usepackage{multirow}
\usepackage{dcolumn}% Align table columns on decimal point
\usepackage{bm}% bold math

\usepackage{times,xspace} 
\usepackage{amsbsy,amssymb,amsmath,bm,bbold} 
\usepackage{graphicx,color,epsfig,rotate} 
\usepackage{fancyhdr} 
\usepackage{epstopdf}
\usepackage{float}

\def\bbbc{{\mathchoice {\setbox0=\hbox{$\displaystyle\rm C$}\hbox{\hbox 
				to0pt{\kern0.4\wd0\vrule height0.9\ht0\hss}\box0}} 
		{\setbox0=\hbox{$\textstyle\rm C$}\hbox{\hbox 
				to0pt{\kern0.4\wd0\vrule height0.9\ht0\hss}\box0}} 
		{\setbox0=\hbox{$\scriptstyle\rm C$}\hbox{\hbox 
				to0pt{\kern0.4\wd0\vrule height0.9\ht0\hss}\box0}} 
		{\setbox0=\hbox{$\scriptscriptstyle\rm C$}\hbox{\hbox 
				to0pt{\kern0.4\wd0\vrule height0.9\ht0\hss}\box0}}}}

\begin{document}

\title{Generation and Stability Analysis of Self Similar Pulses Through Specialty Microstructured Optical Fibers in Mid Infrared Regime}% Force line breaks with \\

\author{Piyali Biswas, Pratik Adhikary, Abhijit Biswas}
\author{Somnath Ghosh}
\email{somiit@rediffmail.com}
\affiliation{Institute of Radiophysics and Electronics\\
 University of Calcutta\\ 92, A.P.C. Road, Kolkata 700009, India}%


%\author{}
%\affiliation{
%\affiliation{
%\affiliation{%


\date{\today}% It is always \today, today,

\begin{abstract}
We report a numerical study on generation and stability of parabolic pulses during their propagation through highly nonlinear specialty optical fibers. Here, we have generated a parabolic pulse at 2.1 $\mu$m wavelength from a Gaussian input pulse with 1.9 ps FWHM and 75 W peak power after travelling through only 20 cm length of a chalcogenide glass based microstructured optical fiber (MOF). Dependence on input pulse shapes towards most efficient conversion into self similar states is reported. The stability in terms of any deviation from dissipative self-similar nature of such pulses has been analyzed by introducing a variable longitudinal loss profile within the spectral loss window of the MOF, and detailed pulse shapes are captured. Moreover, three different dispersion regimes of propagation have been considered to study the suitability to support most stable propagation of the pulse.
%\begin{description}
%\end{description}
\end{abstract}

\keywords{Mid-IR Photonics; Parabolic Pulses; Microstructurd Optical Fibers}%Use showkeys class option if keyword
\maketitle

%\tableofcontents

\section{\label{sec:level1}Introduction}

In mid infrared regime ranging from 2 to 12 $\mu$m high power optical pulses have found their applications in near field microscopy or spectroscopy, mid infrared fiber sources, chemical sensing, biomedical surgeries, imaging and so on \cite{sanghera,ghosh1,sensor}. In this regime, chalcogenide glass based optical fibers are found to be highly efficient in generating high power ultra short optical pulses due to their transperancy to mid IR radiations and extraordinary linear and nonlinear properties \cite{chalco,chalco2,book2}. Parabolic pulses (PP) are a special kind of high power optical pulses that can withstand high nonlinearity of optical fiber being freed from optical wave breaking \cite{wavebreak} and also maintains their parabolic temporal profile throughout the propagation length with their characteristic linear chirp across the pulse width when operated in normal group velocity dispersion (GVD) regime \cite{zhang}. Lately, the generation of high power PPs in fiber amplifiers, fiber bragg gratings and passive fibers have already been demonstrated \cite{fermann,kruglov,parm,barh}. However, most of these studies deliberately ignore any detailed analysis of the self-similar states during propagation through real waveguides. This paper presents a numerical study of generation of a high power parabolic pulse through a chalcogenide glass based optical fiber under various input conditions and analyzes its stability during its propagation through the fiber with high nonlinearity, tailored dispersion and suitably customized losses.

Most of the developments in the generation of parabolic pulses have been done at telecommunication range of wavelengths ($\sim$ 1.55 $\mu$m) and in active media such as fiber amplifiers \cite{wabnitz,finot}. Parabolic pulses have also been generated in a millimeter long tapered silicon photonic nanowire (Si-PhNW) at $\sim$ 2.2 $\mu$m wavelength which is found to be more stable than that generated at 1.55 $\mu$m \cite{lavdas}. Using a chalcogenide glass based microstructured optical fiber (MOF), PP with $\sim$ 4.98 ps full width at half maximum (FWHM) and $\sim$ 46 W peak power has been generated at 2.04 $\mu$m wavelength \cite{barh}. Although these works demonstrate efficient generation of PPs with quite high power and short temporal width at mid IR, less attention has been paid to investigate the stability of such high power pulses. As these optical pulses are extremely short (in the range of picosecond/sub-picosecond), they are highly susceptible to various fiber nonlinearities as well as dispersion behavior which lead them to break after a few centimeters of propagation. Fiber losses with their limited and nonuniform bandwidth are also an important design issue for consideration, as in practice fiber deformations during its fabrication cause losses which turn out to be fatal for an optical pulse. 
In our work, we have generated numerically a parabolic pulse at 2.1 $\mu$m wavelength from an input Gaussian pulse with 75 W peak power and 1.9 ps FWHM after its travel through a 20 cm long arsenic sulphide ($As_2S_3$) matrix based up-tapered MOF. Moreover parabolic pulses generated from different input pulse shapes other than Gaussian have been studied. Accordingly, to study the stability of the generated PP, a variable longitudinal loss profile along with its frequency dependence is incorporated at 2.1 $\mu$m wavelength and the corresponding changes in output pulse characteristics have been reported. Further propagation of the generated PP through different dispersion regimes have been investigated and compared for obtaining the most stable propagation dynamics in such geometries.
\section{\label{sec:level2}Generation of Parabolic Pulse}
\subsection{\label{sec:level3}Numerical Modelling for PP Generation}

The study of most nonlinear effects in optical fibers involves the short pulses with widths ranging from a few picoseconds (ps) to a few femtoseconds (fs). Propagation of such short pulses within the optical fiber is accompanied by dispersion and nonlinearity which influence their shapes and spectra. The pulse evolution along the tapered dispersion decreasing MOF has been modeled by solving the following nonlinear Schr\"{o}dinger equation (NLSE).
Considering a slowly varying pulse envelope $A(z,T)$, NLSE for propagation of short optical pulses takes the form \cite{book},

\begin{equation}
\frac{\partial A}{\partial z}+\beta_1 \frac{\partial A}{\partial T}+i \frac{\beta_2}{2} \frac{\partial^2 A}{\partial T^2} - \frac{\beta_3}{6} \frac{\partial^3 A}{\partial T^3} + \frac{\alpha}{2} A = i \gamma(\omega_0) |A|^2A, 
\end{equation}
where nonlinear parameter $\gamma$ is defined as,
\begin{equation}
\gamma(\omega_0) = \frac{n_2(\omega_0) \omega_0}{c A_e}.
\end{equation}
$\alpha$ is the loss parameter, $\beta_1$ and $\beta_2$ are the first and second order dispersions respectively, $\beta_3$ is the third order dispersion (TOD), $A_e$ is the effective mode area of the fiber and $n_2$ is the nonlinear coefficient of the medium. The pulse amplitude is assumed to be normalized such that $|A|^2$ represents the optical power.
In an ideal loss-less optical fiber with normal GVD i.e., when value of $\beta_2$ is positive and a hyperbolic dispersion decreasing profile along length of the fiber, the asymptotic solution of NLSE yields a parabolic intensity profile. Under this condition, the propagation of optical pulses is governed by the NLSE of the form \cite{zhang},

\begin{equation}
\label{eq:3}
i \frac{\partial A}{\partial z}-\frac{\beta_2}{2} D(z) \frac{\partial^2 A}{\partial T^2}-i \frac{\beta_3}{6} \frac{\partial^3 A}{\partial T^3}+ \gamma(z)|A|^2A = 0, 
\end{equation}
where $D(z)$ is length dependent dispersion profile along the tapered length, $\beta_2$($2^{nd}$ order GVD parameter)$>0$, $\beta_3$ is the TOD value and $\gamma(z)$ is longitudinally varying nonlinear (NL) coefficient. By making use of the coordinate transformation, $\xi=\int_{0}^{z}D(z')dz'$ and defining a new amplitude $U(\xi,T)=\frac{A(\xi,T)}{\sqrt{D(\xi)}}$, eq. (3) transforms to,

\begin{equation}
\label{eq:4}
i \frac{\partial U}{\partial \xi}-\frac{\beta_2}{2} \frac{\partial^2 U}{\partial T^2}-i \frac{\beta_3}{6 D(\xi)}\frac{\partial^3 U}{\partial T^3}+ \gamma(z)|U|^2U = i \frac{\Gamma(\xi)}{2}U, 
\end{equation}
where \begin{equation} 
\label{eq:5}
\Gamma(\xi)=-\frac{1}{D}\frac{dD}{d\xi}=-\frac{1}{D^2}\frac{dD}{dz}
\end{equation}

As $D(z)$ is a decreasing function of $z$, $\Gamma$ in eq.(\ref{eq:5}) is positive since $D$ is a decreasing function with increasing $z$; and hence it mimics as a gain term in eq.(\ref{eq:3}). In the chosen dispersion decreasing fiber (DDF), the varying dispersion term is equivalent to the varying gain term of a fiber amplifier with normal GVD. Specifically, with the choice of 
$D(z) = \frac{1}{1+\Gamma_0 z}$ the gain coefficient becomes constant, i.e., $\Gamma$ = $\Gamma_0$.
The NLS equation in a fiber with normal GVD and a constant gain coefficient permits self-similar propagation of a linearly chirped parabolic pulse as an asymptotic solution.

To study the pulse propagation in nonlinear dispersive media Split-step Fourier method (SSFM) has been extensively efficient, which is much faster than any other numerical approach to achieve the same accuracy. In general, dispersion and nonlinearity act together along the length of the fiber. SSFM obtains an approximate solution by assuming that in propagating the optical field over a small distance $h$, the dispersion and nonlinear effects can be considered to act independently. More specifically, propagation from $z$ to $z+h$ is carried out in two steps. In the first step, nonlinearity acts alone while in the second step dispersion acts alone.  In this method eq.(\ref{eq:4}) can be written in the form \cite{book}

\begin{equation}
\frac{\partial A}{\partial z}=(\hat D+\hat N)A,
\end{equation}
where $\hat D$ is the differential operator that accounts for the dispersion and losses within the medium and $\hat N$ is the nonlinear operator that governs the effect of fiber nonlinearities on pulse propagation. 

\subsection{Pulse Evolution}
\begin{figure}[htbp]
	%\centering
	\includegraphics[width=8cm]{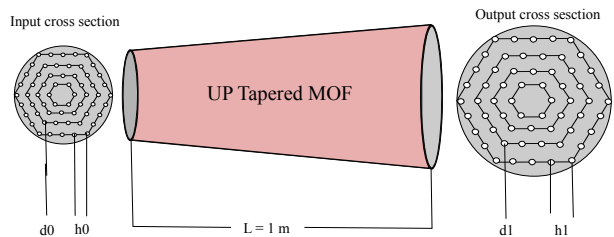}
	\caption{(color online) A schematic of the linearly up-tapered MOF. The length of the MOF is considered to be 1 m. At the input end d0 is the individual air hole diameter and d1 the same at the output end, h0 and h1 are the air hole separation at the input and output end, respectively. The chosen taper ratio is 1.05}
	\label{fig:figure1}
\end{figure}

\begin{figure}[htbp]
	%\centering
	\includegraphics[width=8cm]{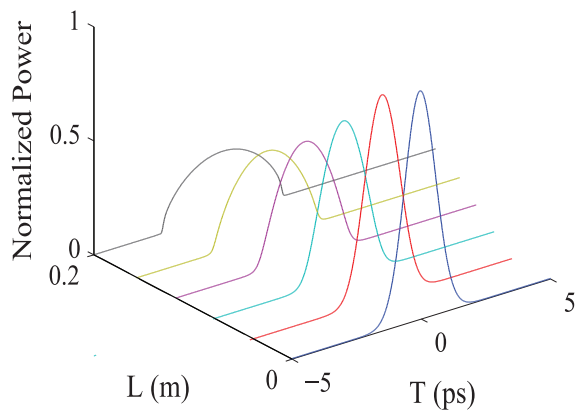}
	\caption{(color online) Pulse evolution from Gaussian input pulse (blue curve) to parabolic pulse (black curve) at only 20 cm length of the fiber.}
	\label{fig:figure2}
\end{figure}

\begin{figure}[htbp]
	%\centering
		\includegraphics[width=8cm]{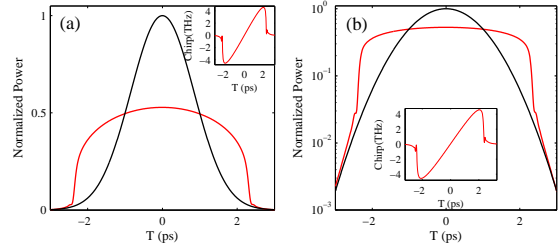}
		\caption{(color online) (a) Time domain plot of the input Gaussian (black) and the output parabolic pulse (red). The linear chirp across the parabolic pulse width is shown in the inset, (b) the logarithmic plot of the input (black) and the output (red) pulses with linear chirp profile shown in the inset.}
	\label{fig:figure3}
\end{figure}

We aim to generate a high power parabolic pulse in the mid infrared regime. A parabolic pulse has been efficiently generated through numerical simulation at 2.1 $\mu$m wavelength in arsenic sulphide ($As_2S_3$) based MOF geometry with a solid core, surrounded by a holey cladding consisted of 4 hexagonally arranged rings of air holes embedded in the $As_2S_3$ matrix \cite{ghosh1,barh}. $As_2S_3$ possesses lowest transmission loss ($\alpha_T$ $\sim$ 0.4 dB/m at 2 $\mu$m) among chalcogenide glasses and very high nonlinearity ($n_2$ $\sim$ $4.2\times10^{-18}$ $m^2$/W at 2 $\mu$m). A meter long up-tapered MOF with suitably tailored dispersion and nonlinearity is shown in figure \ref{fig:figure1}. A Gaussian pulse of peak power 75 W and initial full-width-at-half-maximum (FWHM) 1.9 ps was fed at the input end of the fiber and after propagating only 20 cm of the fiber length, the shape of the pulse in time domain is transformed into parabolic. The pulse evolution is shown in figure \ref{fig:figure2}. The Gaussian pulse is reshaped to parabolic under the combined influence of self phase modulation (SPM) and normal GVD. In figure \ref{fig:figure3}(b), the top-hat nature of the output pulse in logarithmic scale carries the hallmark that the generated pulse is essentially parabolic. With the inclusion of third order dispersion (TOD), the parabolic profile of the pulse is still maintained. The output pulse is broadened up to 4.65 ps (figure \ref{fig:figure3}(a)) and a linear chirp is generated across the entire pulse width as shown in the inset of figure \ref{fig:figure3}(a).


\subsection{Effect of Input Pulse Shapes}
\begin{figure}[htbp]
	%\centering
		\includegraphics[width=8cm]{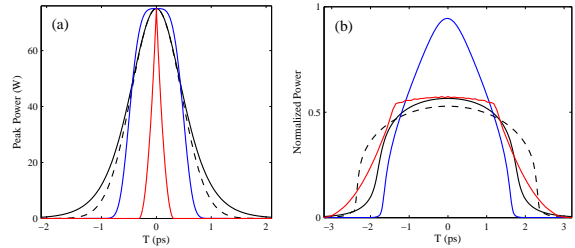}
		\caption{(color online) Plot of (a) four different input pulse shapes - Gaussian (dashed black), hyperbolic secant (solid black), triangular (red) and supergaussian (blue), all having the same energy 159 pJ; and (b) output pulses obtained from various input pulses.}
	\label{fig:figure4}
\end{figure}

In this section we will discuss how various input pulse shapes affect the parabolic pulse evolution through the MOF. In order to investigate this we employ different pulse shapes at the input, such as a hyperbolic secant pulse with $A(0,T) \propto  sech(T/T_0)$, a triangular pulse with $A(0,T) \propto  tripuls(T/T_0)$ and a supergaussian pulse with $A(0,T) \propto  \exp(-(T/T_0)^{2m})$. The input pulses are unchirped with the same peak power and energy as shown in figure \ref{fig:figure4}(a) and the medium of propagation is assumed to be lossless. 
First, we consider a secant hyperbolic pulse with 159 pJ of energy and 1.685 ps FWHM which after propagating through 15 cm of the fiber length is converted to a parabolic intensity profile having a temporal FWHM of 3.48 ps. The output spectral broadening is estimated to 125 nm. Relative to the PP evolved from a Gaussian input, corresponding PP for the counterpart secant pulse in both temporal and spectral domain is less broadened. 

With a triangular pulse of energy 159 pJ and FWHM of 1.70 ps, a nearly parabolic pulse is generated after 17 cm propagation through the fiber. Spectral output is broadened by 110 nm.

Further, a supergaussian pulse of the same energy and FWHM of 2.21 ps is fed at the input end of the fiber. After propagating only 8 cm, a very weak parabolic intensity profile with FWHM 2.38 ps and a nearly linear chirp across is observed. Further propagation of the pulse results in a triangular shaped temporal profile. The output pulse shapes obtained at optimum length of the fiber are depicted in figure \ref{fig:figure4}(b). A comparative study for the characteristic parameters of the output pulses for different input pulse shapes are presented in Table~\ref{jlab1}.

\begin{table*}[t]
	\centering
	\caption{\label{jlab1}Comparison of output pulses generated from various input pulse shapes.}
	\footnotesize
	%\begin{indented}
	\begin{tabular}{|c|c|c|c|c|c|c|c|c|}
		\hline
		Input&Input&Input&Optimum fiber&Fiber length&Output&Output&Energy&Spectral\\
		pulse&energy&FWHM&length for PP&before wave&FWHM&energy&conversion&broadening\\
		shapes&(pJ)&(ps)&generation (cm)&breaking (cm)&(ps)&(pJ)&efficiency (\%)&(nm)\\
		\hline
		Gaussian&159&1.90&15&25&4.05&159&100&122\\
		Hyperbolic&159&1.52&15&22&3.48&159&100&125\\
		secant&&&&&&&&\\
		Triangular&159&1.70&17&19&3.72&159&100&110\\
		Supergaussian&159&2.21&8&13&2.38&159&100&120\\
		\hline
	\end{tabular}\\
	%\end{indented}
\end{table*}
\normalsize

\section{Stability of Similariton Propagation}
\subsection{Loss Window}
We start our study with the generation of parabolic pulses in a lossless, suitably dispersion and nonlinearity tailored fiber. Here we examine the stability of the generated PP under the influence of a lossy medium.

\begin{figure}[h]
	%\centering
		\includegraphics[width=8cm]{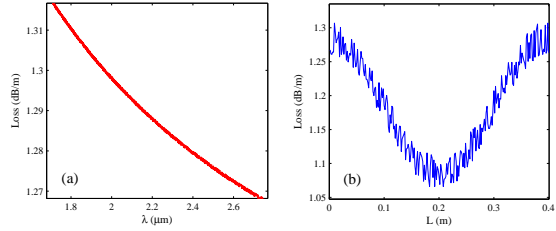}
		\caption{(a) Loss window for $As_2S_3$ matrix based chalcogenide MOF and (b) loss variation of the fiber along the fiber length.}
	\label{fig:figure5}
\end{figure}

For this purpose we considered the specific material loss window for the $As_2S_3$ chalcogenide MOF around 2.1 $\mu$m wavelengths, following \cite{losswindow}, as shown in figure \ref{fig:figure5}(a). The material loss of $As_2S_3$ glass is around 0.3 dB/m \cite{losswindow} and the confinement loss of the MOF is taken as 1.0 dB/m \cite{barh}. So a total loss of 1.30 dB/m has been considered as the mean value and certain amount of deliberate fluctuations is introduced. Additionally, our investigation include two different mean values of the over all loss exceeding the previous value.

\subsection{Loss Fluctuations}
\begin{figure}[htbp]
	%\centering
		\includegraphics[width=8cm]{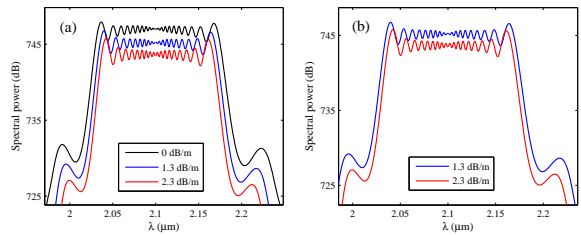}
		\caption{(color online) Various output spectra obtained at different loss values with (a) 10\% and (b) 20\% fluctuations of the longitudinal loss profiles respectively.}
	\label{fig:figure6}
\end{figure}

In order to check the stability of the output pulse spectrum, we introduce a variable loss along the fiber length with certain amount of randomness as shown in figure \ref{fig:figure5}(b), corresponding to three different loss values. To address the tolerance issue of the state-of-the-art fabrication process in terms of loss variation along the fiber length, loss fluctuations as high as 10\% and 20\% around the mean values have been considered. Figure \ref{fig:figure6}(a) illustrates the spectral power reduction due to 10\% fluctuations of various loss values as compared to the lossless spectrum. Accordingly, spectral power change due to 20\% fluctuations of loss values are shown in figure \ref{fig:figure6}(b). Exact quantifications of the spectral modifications in terms of 3dB bandwidth are shown in Table \ref{jlab2}.

\begin{table}[htbp]
	\centering
	\caption{\label{jlab2}Comparison of PP with variable loss effect.}
	\footnotesize
	%\begin{indented}
	\begin{tabular}{|c|c|c|c|c|}
		\hline
		Loss&Loss&3dB&Maximum&Output\\
		(dB/m)&fluctuation&Bandwidth& spectral&energy (pJ)\\
		&(\%)&change (\%)&rippling (dB)&\\
		\hline
		0&0.0&0.0&2.5&159\\
		\hline
		\multirow{2}{*}{1.30}&10&5.5&2.9&127\\
		&20&6.9&3.0&\\
		\hline
		\multirow{2}{*}{1.80}&10&8.3&3.4&116\\
		&20&9.8&3.5&\\
		\hline
		\multirow{2}{*}{2.30}&10&8.9&3.5&105\\
		&20&9.6&3.6&\\
		\hline
	\end{tabular}\\
\end{table}

\subsection{Different Dispersion Regimes}
\begin{figure}[htbp]
	\centering
	\includegraphics[width=8cm]{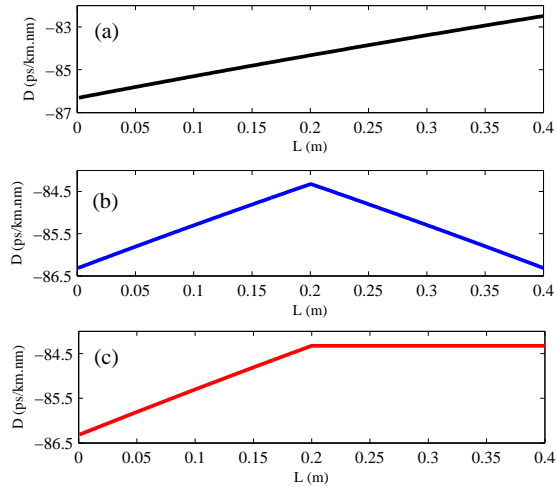}
	\caption{Dispersion profiles of the (a) up-taper MOF structure, (b) up-down taper MOF and (c) up-no taper MOF.}
	\label{fig:figure7}
\end{figure}

For the stability analysis of the generated parabolic pulse, we use a two stage propagation of the pulse. Once a parabolic pulse is generated, it has been shown in various experiments that it retains its shape throughout the propagation length and follows self-similar propagation. To investigate the self similar propagation of the PP through a passive medium, we consider MOFs with three different configurations in which the parabolic pulse will be generated in first few centimeters of the fiber length. Hence the PP will be propagating through rest of the fiber length engineered with three distinct dipersion profiles, respectively. Chosen relevant fiber geometries are shown in figure \ref{fig:figure7}.

\begin{figure}[htbp]
	%\centering
		\includegraphics[width=8cm]{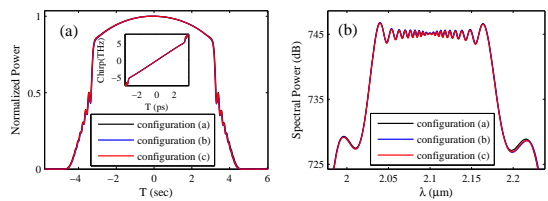}
		\caption{(color online) (a) Plot of temporal profiles of the output pulses obtained from three different MOF configurations at the end of 40 cm length and (b) spectral variation of the corresponding output pulses.}
	\label{fig:figure8}
\end{figure}

Firstly, we consider a fully up tapered MOF in which, up to 20 cm from the input end of the MOF, evolution of the parabolic pulse from a Gaussian seed pulse has been observed. Through rest of the fiber length, the self similar characteristic of the generated PP has been studied. Before explaining the obtained results from this up-tapered MOF, we will reconsider this pulse evolution process in other two cases. The second kind of fiber geometry under consideration is an up-down tapered fiber. Here, the first 20 cm of the total fiber is a dispersion decreasing MOF to genearte the PP efficiently. Then this PP has been fed to a down tapered MOF of same material as the up tapered fiber with increasing dispersion profile along the fiber length. Finally, the parabolic pulse generation and propagation have been studied in an up-straight MOF, where the PP is evolved through 20 cm fiber length and its propagation is made through a untapered MOF with a constant dispersion profile. Results of these chosen three different configurations are shown in figure \ref{fig:figure8}.

The output pulses and spectra after propagation through 40 cm fiber length respectively, look almost identical irrespective of their fiber geometries. Notably the striking feature is that the pulses are no longer parabolic in shape as we could expect from the self similar propagation characteristics of parabolic pulses. Rather, we have obtained a parabolic pulse unlike a similariton. The input pulse has evolved to a parabolic shape at 20 cm of the fiber length which is just a transient state of the input pulse evolution in the passive media. On further propagation it essentially transformed into a nearly trapezoidal shape with linear chirp across most of the pulse. As our chosen MOF is a highly nonlinear fiber (HNLF), the large $\gamma$/$\beta_2$ ratio makes SPM to dominate over dispersion. The combined effect of SPM and GVD has resulted in broadening of the pulse but failed to maintain its shape. If we examine the chirp variation of the parabolic pulse in figure \ref{fig:figure3}, it could be seen that its linear nature extends almost over the entire pulse width with steepened transitions at the leading and trailing edges. However in all three chosen fiber configurations (as shown in figure \ref{fig:figure7}) the chirp evolution of the propagating pulse carries an interesting signature of flipping the both edges in temporal domain. In addition as anticipated, the parabolic profile is gradually transformed into a nearly trapezoidal profile with increasing propagation length. The nonmonotonic nature of the chirp is somewhat responsible for the re-reshaping of the propagating pulse \cite{nfactor}. The output spectra carry signature of spectral broadening up to 140 nm and a nearly flat top (fluctuations falls within 3 dB). The unwanted side side-lobes appears as a result of interference between newly generated frequency components.

\section{Conclusion}
In conclusion, numerically generated parabolic pulses from various input optical pulse shapes such as Gaussian, hyperbolic secant, triangular and super-gaussian keeping initial energy constant has been demonstrated. A in-depth qualitative and quantitative analysis of these PPs has established that the PP obtained by reshaping of the input Gaussian pulse has turned out to be the most efficient irrespective of the choice of chalcogenide glass based MOF designs. Moreover, the PP generation in different length dependent dispersion regimes such as up-taper, up-down taper and up-no taper geometries has been studied. From a direct comparison of PPs obtained from different structures, its evident that PPs look almost similar in every aspect for different cases. Hence, from practical point of view we may propose the up-tapered MOF geometry as preferable fiber structure for generating PP owing to its fabrication friendly geometry. In addition, the stability of the PP has been investigated by introducing longitudinally variable and customized fluctuating loss profiles within the specific loss window of the chosen MOFs. From the pulse dynamics through these dissipative structures, though no significantly adverse effect particularly on the shape of the output spectra  was observed, however a reduction in the spectral power along with lesser 3dB bandwidth has been noticed. Moreover, from the propagation characteristics through the dispersion tailored MOFs, it has been settled that the generated parabolic shape is a transient state which merely capable of retaining its shape unless an optimized fiber design/scheme is proposed for self-consistent solution. Our findings would be of key interest for design and fabrication of self-consistent and stable PP sources for mid infrared spectroscopy, fiber-based biomedical surgeries, chemical sensing etc.     

\section*{Acknowlegdement}
S. N. Ghosh acknowledges the financial support by Department of Science and Technology (DST), India as a INSPIRE Faculty Fellow [IFA-12;PH-13].

%\section*{References}

\end{document}